\documentclass[grl]{agu2001}

\usepackage{epsfig}

\authorrunninghead{USOSKIN ET AL.}
\titlerunninghead{ LOW CLOUDS VS. COSMIC RAY INDUCED IONIZATION}

\journalid{XXX 2003} \articleid{XXX}{AX} \paperid{2004GL019507}
\cpright{AGU}{2003}

\received{Month XX, 2003} \revised{Month XX, 2003}
\accepted{Month XX, 2003} \published{}

\authoraddr{O.~G. Gladysheva and G.~A. Kovaltsov,
Ioffe Physical-Technical Institute, St.Petersburg, Russia}

\authoraddr{N. Marsh,
Danish Space Research Institute, Copenhagen, Denmark}

\authoraddr{K. Mursula,
Department of Physical Sciences, FIN-90014 University
of Oulu, Finland}

\authoraddr{I.~G. Usoskin,
Sodankyl\"a Geophysical Observatory (Oulu unit),
FIN-90014 University of Oulu, Finland, (Ilya.Usoskin@oulu.fi)}

\begin{document}

\title{Latitudinal dependence of low cloud amount on cosmic ray induced ionization}

\author{I.G.   Usoskin, N.~Marsh, G.A. Kovaltsov, K.~Mursula, O.G. Gladysheva}


\begin{abstract}
A  significant correlation  between the  annual cosmic ray flux and
 the amount of  low clouds has  recently been found  for the
 past 20  years.
However, of the physical explanations suggested, none has been
 quantitatively verified in the atmosphere by a combination of modelling and experiment.
Here we study  the  relation  between the global distributions of
 the  observed low cloud  amount and the calculated tropospheric ionization induced by cosmic rays.
We find that the time evolution of the low cloud amount can be decomposed into a
 long-term trend and inter-annual variations, the latter depicting a clear 11-year cycle.
We also find that the relative inter-annual variability in low cloud amount increases polewards
 and exhibits a highly significant one-to-one relation with inter-annual variations in
 the ionization over the latitude range 20--55$^\circ$S and  10--70$^\circ$N.
This latitudinal dependence gives strong support for the hypothesis that the
 cosmic ray induced ionization modulates cloud  properties.
 \end{abstract}

\begin{article}
\section{Introduction}

\begin{figure*}
\begin{center}
\resizebox{15cm}{3.5cm}{\includegraphics{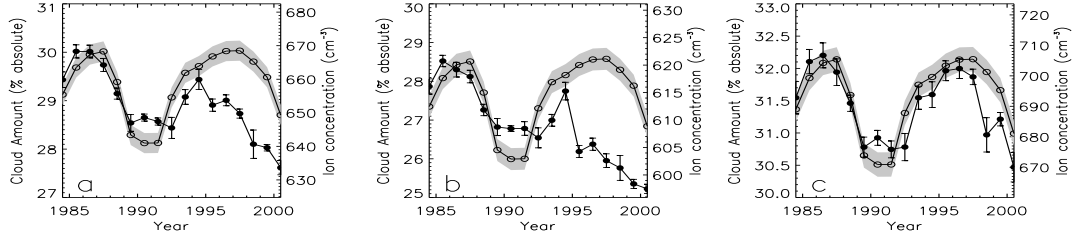}}
\end{center}
\caption{Time profiles of LCA in percent of the area coverage
     (solid  symbols, left  axis)  and CRII
 (open  symbols,  right axis)  for  a) the  global average  ($60^{\circ}S
 <\lambda < 70^{\circ}N$), b) tropics ($|\lambda| < 25^{\circ}$), and c) mid-latitudes  ($60^{\circ}S < \lambda  < 25^{\circ}S$
 and $25^{\circ}N <\lambda < 70^{\circ}N$). Error bars for LCA ($\pm \sigma$)
 are estimated for each annual average from the corresponding monthly fluctuations,
 after removal of the seasonal cycle.
 Any effects due to instrument or calibration uncertainties are neglected.
 Errors in CRII ($\pm \sigma$, grey shading) correspond to $\pm$50 MV
 uncertainties in the reconstructed annual modulation strength \citep{usos02}.}
     \label{Fig:1_full}
\end{figure*}
A  possible influence of  solar variability on climate  has been
 discussed for some time.
Although the direct solar influence on climate  is apparent,  variations
 of  the solar irradiance are estimated  to  be an  order of
 magnitude  too  small  to explain  the observed   changes  in  climate
 \cite[e.g.,][]{stot03}.
Therefore, an indirect mechanism linking solar variability to climate should be  involved.
According to  some modeling studies,  a  response  in  atmospheric  circulation  can  amplify  the
terrestrial  effect of solar  irradiance changes  \citep{Haigh02}.
On the other  hand, cosmic
 rays can noticeably affect the Earth's climate \citep{sven97,mars00,cars02,shaviv03}.
While  the energy deposited by cosmic rays into the Earth's atmosphere is negligible compared to that
from solar irradiance,  they are the main source  of ionization in the
troposphere \citep[see, e.g.][]{Baz00}.

A possible qualitative link has been proposed that relates cosmic ray
 induced ionization (CRII) in the troposphere and cloud  properties \citep{sven97,mars00}.
Ions created by cosmic rays rapidly interact with molecules in the
 atmosphere and are converted to complex cluster ions (aerosols) \citep{gring86,Hopp86}.
These cluster ions may grow by ion-ion recombination or ion-aerosol
 attachment and thus affect the number of aerosols acting as cloud condensation nuclei
 (CCN) at typical atmospheric supersaturations of a few percent \citep{vigg95,yu01}.
Others have suggested that a CRII-cloud link could also arise through changes in the
 global electric circuit affecting aerosol-cloud interactions at the edges of clouds
 \citep[see, e.g.,][or a review of possible mecahnisms in {\it Harrison and
 Carslaw}, 2003]{Tinsley00}.
Both mechanisms require that an amplified effect
of cosmic rays on climate is realized through the important role
that clouds  play in  the radiation budget  of the atmosphere  by both
trapping outgoing  long wave  radiation and reflecting  incoming solar
radiation.   Although  a  detailed  physical  model  quantifying  this
connection is still missing, correlation studies support its validity.
\citet{mars00}  found  a highly  significant  correlation between  low
clouds below $\sim$ 3.2km (rather than clouds at other altitudes) and the cosmic ray flux during
the  period  1983--1994.   This  basic result  has  subsequently  been
confirmed by  other independent studies  \citep{pall00,yu02}.
There is  also evidence for the  reduction of cloud coverage  during strong
Forbush decreases  at time scales  of a few days  \citep{pudo96}. This
implies that the proposed cloud-cosmic ray relation may also be significant at
short-time scales.
More recently \citet{mars03} found that the low cloud-cosmic ray correlation can be extended until 2001
 but only after the globally averaged cloud data are re-calibrated.
However, the variability in low cloud amount (LCA) cannot be uniquely ascribed to a single
 mechanism when using globally averaged data since the observed long-term changes
 in the global LCA correlate with different solar-related indices including
 solar irradiance and cosmic rays.

In this paper we study the spatial distribution of LCA and CRII over the period 1984--2000.
In all previous studies the count rate of a single neutron monitor was used
 as a measure of cosmic rays, and assumed to represent the global
 CRII$^1$\footnotetext{$^1$ After submission of this paper we were made aware of the Ph.D.
 thesis by E. Pall\'e (The Queens University of Belfast, 2001) where CRII was also calculated.
 However, none of the main conclusions of this paper were obtained or discussed by Pall\'e.}.
Although useful for qualitative correlation studies, this approach does not give
 quantitative estimates since the cosmic ray intensity varies strongly
 over the globe due to the shielding by the geomagnetic field.
Here we study the global distribution of CRII and compare that with the
 measured LCA distribution.

\section{LCA-CRII relations during 1984--2000}

Following previous studies \citep{mars03}, we use the low cloud amount obtained from the
 ISCCP-D2 database limited to IR radiances. ISCCP provides monthly observations of  the global
 cloud cover based on an intercalibration of up to 5 satellites for the
 period from July 1983 to September 2001.
Satellites detect a cloud when radiance observations differ significantly from clear sky values.
However, uncertainties can arise if atmospheric transparency is influenced by processes
 other than clouds, e.g., aerosol loading from Mt. Pinatubo \citep{Luo02}.
We note that LCA as defined from satellite observations is restricted to clouds with
 their tops below 640 hPa (3.2km), which is different from ground-based observations.
In the  present analysis  annual LCA averages are  used (in order to avoid
 seasonal variations) on a 5$^{\circ}$x5$^{\circ}$ latitude-longitude grid
 for the period 1984--2000 inclusive.

Recently, the global distribution of CRII has been calculated for the
 troposphere (0--10 km) since 1951 \citep{usos03}.
First, the electromagnetic-nucleonic cascade initiated by cosmic rays in the atmosphere
 was simulated for different conditions using the CORSIKA Monte-Carlo package
 \citep{heck98}.
Then the annually averaged ion production rate in the troposphere at a given latitude was
 calculated using the respective cosmic ray spectra parameterized by the average
 heliospheric modulation strength \citep{usos02}.
Finally, the equilibrium ion concentration was calculated at a given location, taking
 into account processes of recombination and aerosol attachment.
Here we use CRII values calculated at 3 km altitude which corresponds
 roughly to the limiting altitude, as defined by ISCCP-D2, below which low cloud forms.

\begin{table}
\begin{tabular}{p{2.5cm}|ccc}
\hline\hline Data & global & tropics & mid-latitudes\\
\hline Raw data
&  0.46 (61\%)  & 0.14 (26\%) &  0.81 (98\%)  \\
De-trended data  & 0.84 ($>$99\%) &  0.61 (94\%)  & 0.90 ($>$99\%)  \\
\hline\hline
\end{tabular}
\caption{Correlation coefficients (and their significance levels in parentheses)
 between LCA and CRII for the period of 1984--2000
 for different regions:  global ($60^{\circ}S <\lambda <70^{\circ}N$),
 tropics  ($|\lambda|< 25^{\circ}$),  and  mid-latitudes ($60^{\circ}S
 <\lambda  <25^{\circ}S$ and $25^{\circ}N  < \lambda  < 70^{\circ}N$).}
 \label{Tab}
\end{table}

Time profiles of measured LCA and calculated CRII are shown in Fig.~\ref{Fig:1_full}
 for different regions, the corresponding values of the correlation
 coefficient (c.c.) and their significance levels are summarized in the first
 row of Table~\ref{Tab}.
Polar regions ($\lambda>60^\circ$S and $\lambda>70^\circ$N) are excluded from the analysis
 in order to avoid the problems associated with cloud detection over ice.
The rest of the globe (60$^\circ$S$ < \lambda < 70^\circ$N) is further divided
 into two latitudinal regions: tropics ($|\lambda|<25^\circ$) and middle latitudes ($\lambda=[25^\circ -60^\circ]$S
 and $[25^\circ - 70^\circ]$N).
Similar to previous studies \citep{mars03}, the statistical significance of c.c.
 has been estimated using the random phase test \citep{ebis97}.

\begin{figure}
 \begin{center}
\resizebox{!}{3cm}{\includegraphics{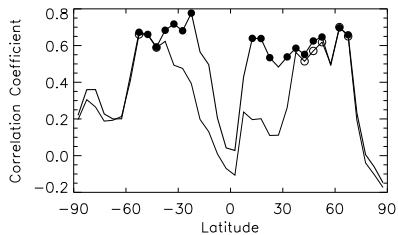}}
 \end{center}
\caption{Latitudinal dependence of the cross-correlation coefficient between LCA
      and CRII for 1984--2000.  Results for raw and detrended LCA data are shown
      by thin and thick lines, respectively. Correlation coefficients above $90\%$
      significance level are indicated with symbols.}
     \label{Fig:2_zonal}
\end{figure}
\begin{figure}
\resizebox{8cm}{!}{\includegraphics{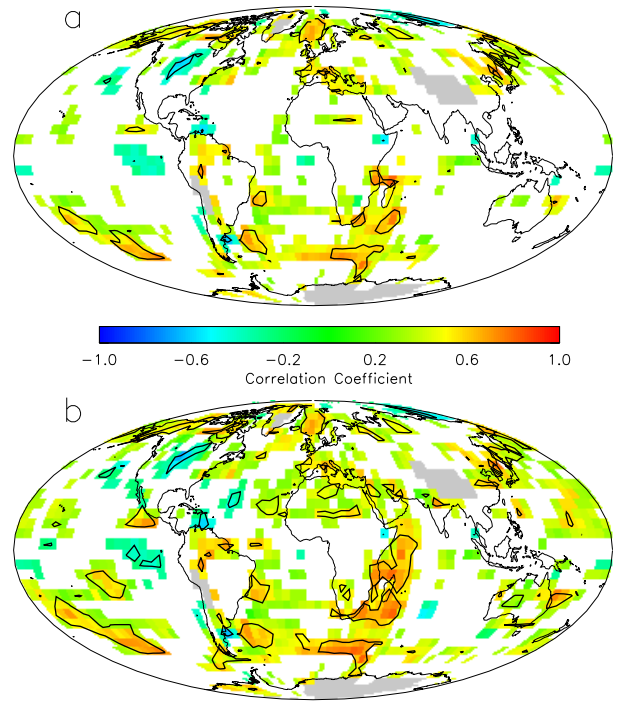}}
\caption{Global distribution of the correlation coefficients between CRII
       and  LCA for  1984--2000 using the  raw (panel  a) and
detrended (panel b) cloud data.   Only
areas  with significant correlation  (significance level  $>68\%$) are
shown while areas  of highly  significant correlation  ($>90\%$) are
 indicated by the thick contour line. Areas with no data are given in grey.}
     \label{Fig:3_corr}
\end{figure}

The c.c. between zonal averages (within 5$^{\circ}$
 latitudinal  belts) of CRII and LCA are  depicted  by  the  thin line  in
 Fig.~\ref{Fig:2_zonal}.
The global distribution of the significant c.c. within a 5$^\circ$x5$^\circ$ grid
 is shown in Fig.~\ref{Fig:3_corr}a.
One can see that the significant coefficients are not uniformly  distributed over the
 globe.
The correlation is high at middle latitudes but is suppressed in tropical
 regions, leading to a moderate global correlation (see also Fig.~\ref{Fig:2_zonal}).
A similar conclusion has been drawn for total clouds by \citet{sven97} and for
 low clouds by \citet{mars03} who suggested that ENSO dominates inter-annual variability in the tropics.

LCA and CRII behave very similarly to  each other at middle
 latitudes, both depicting the dominant 11-year cycle (Fig.~\ref{Fig:1_full}c),
 but are somewhat  different in  the tropics.
While the CRII time series has qualitatively the same form in all geographical zones,
 LCA  behaves differently.
A strongly decreasing trend of about 0.2\% per year is apparent in LCA time profile in
 the tropical regions (Fig.~\ref{Fig:1_full}b) onto which an 11-year cycle
 is superimposed.
The trend is also clearly seen in the global LCA average (Fig.~\ref{Fig:1_full}a)
 \citep[cf.,][]{mars03},
  while the corresponding trend in CRII is close to zero ($0.2\pm0.5$ cm$^{-3}$/year).
The trend is not uniform over the globe --
 while the trend is mostly weak in mid-latitude regions, tropical  regions are
 dominated by  areas of strong decreasing trend.
Such a trend can mask the agreement between the variations of
 LCA and CRII during the period 1984-2000.
Accordingly, LCA can be decomposed into a long-term trend and shorter-term
 inter-annual variations around this trend.
The origin of this trend could be related to physical processes, e.g., a change in the global circulation
 pattern or an increased loading of atmospheric aerosol, or to an instrumental effect, e.g.,
 the inter-calibration of satellites providing global cloud observations as suggested
 by \citet{mars03}.
In the following only the detrended inter-annual variations of LCA are considered.
We suggest that CRII is not the main source of cloud formation but rather "modulates" it,
 and that the long-term trend results from other processes, which are outside
 the main focus of this study.

\begin{figure*}
 \begin{center}
\resizebox{15cm}{3.5cm}{\includegraphics{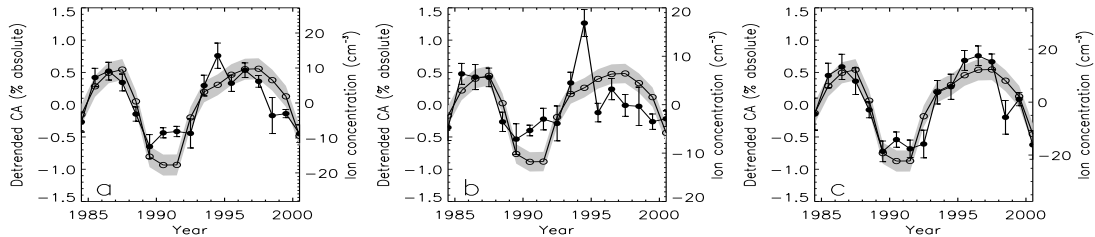}}
 \end{center}
\caption{The same as in Fig.~\ref{Fig:1_full} but for the detrended variations of LCA and CRII.}
     \label{Fig:5_detrended}
\end{figure*}

Using a linear approximation for the long-term trend during
 1984--2000, $\overline{LCA}(t)=LCA_o + B\cdot t$, we have investigated the
 detrended variations of LCA, $\Delta LCA  \equiv LCA-\overline{LCA}$.
Fig.~\ref{Fig:3_corr}b shows the spatial distribution of c.c. between $\Delta  LCA$ and CRII.
While the total area of significantly negative c.c. is very small and is not greatly affected
 by detrending LCA, areas of significantly positive c.c. occupy a large
 fraction of surface covered by the cloud data (see Table~\ref{Tab:area}).
We note that areas of significantly positive c.c. show a tendency towards the geographical pattern
 reminiscent to that of warm ocean currents flowing from the equator towards
 higher latitudes.
Comparing  the two maps in Fig.~\ref{Fig:3_corr} one can see that detrending
 LCA greatly improves the correlation with CRII (see also Tables \ref{Tab} and \ref{Tab:area}).
This is also clearly seen from the correlation between zonal averages of $\Delta   LCA$,
 and CRII represented by the thick line in Fig~\ref{Fig:2_zonal}.
Due to detrending of LCA data, the latitude range possessing highly significant
 positive c.c. (s.l.$>$90\%) is increased from  [50--55$^\circ$S;
 40--70$^\circ$N] to [20--55$^\circ$S;  10--70$^\circ$N] (see Fig.~\ref{Fig:2_zonal}).
Fig.~\ref{Fig:5_detrended} illustrates that there is now similarity between $\Delta$LCA
 and CRII time profiles all over the globe (excluding polar regions).

\begin{table}
\begin{tabular}{c|cccc}
\hline\hline   Data  &signif.&high  signif.&   signif.&high  signif.\\
&negative&negative&positive&positive\\ \hline  Raw data & 4\%  & 1\% &
25\%  & 6.5\%  \\ De-trended  data &  4.5\%  & 1\%  & 39\%  & 15\%  \\
\hline\hline
\end{tabular}
\caption{Fraction of the global surface (areas with no cloud data  are  excluded)
 covered  by  significant  (s.l.$>$68\%)  and  highly
significant   (s.l.$>$90\%)   correlation   (positive   and   negative
separately) between CRII and raw and detrended LCA data.}
 \label{Tab:area}
\end{table}

\begin{figure}[t]
\begin{center}
\resizebox{8cm}{!}{\includegraphics{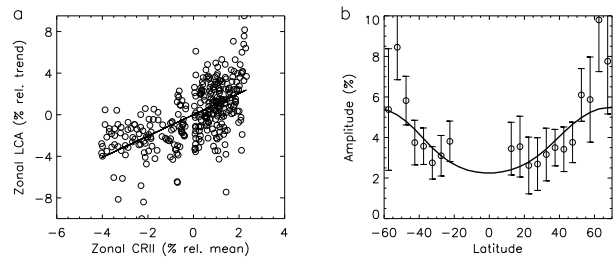}}
\end{center}
\caption{Latitudinal relation between relative variations of $\delta (LCA)$ and $\delta (CRII)$
  for the period 1984--2000 within the latitude range $55^{\circ}-20^{\circ}$S and $10^{\circ}-70^{\circ}$N.
  a) Scatter plot of $\delta (LCA)$ vs. $\delta (CRII)$, each dot representing an annual value within a
    $5^\circ$ latitudinal bin. Solid line depicts the best linear fit (Eq.~\ref{Eq:lin2}).
  b) Latitudinal dependence of the amplitude of cyclic variations in $\delta (LCA)$ (dots) and
   $\delta (CRII)$ (line). The amplitude is found by fitting a 10-year sinusoid to the respective time profiles. }
     \label{Fig:6_scatter}
\end{figure}

Using  the latitudinal zones with highly significant c.c.'s (see Fig.~\ref{Fig:2_zonal}) we
 try to quantify the relation between de\-trended LCA and  CRII (Fig.~\ref{Fig:5_detrended}).
For want of a physical model relating LCA to CRII, a quantitative phenomenological
 relation is assumed in the form of a direct proportionality between
 normalized variations of LCA, $\delta (LCA)=\Delta LCA/\overline{LCA}$, and
 CRII, $\delta (CRII)=(CRII-\overline{CRII})/\overline{CRII}$, where
 $\overline{CRII}$ is the zonal mean CRII value during 1984--2000.
The scatter plot of $\delta (LCA)$ vs.  $\delta (CRII)$   is  shown   in
 Fig.~\ref{Fig:6_scatter}a.
Despite  the wide  scatter of  points, there  is a  highly significant
 correlation  between  $\delta (LCA)$  and  $\delta (CRII)$  (c.c.=0.6,
 s.l.$>$98\%), with the corresponding linear relation as follows:
\begin{equation}
\delta (LCA) = (1.02\pm 0.08) \delta (CRII). \label{Eq:lin2}
\end{equation}
The  fact  that the  proportionality  coefficient  is  close to  unity
 implies  that inter-annual  variations  of LCA  around the  long-term
 trend can be directly ascribed to  the variations of CRII.
Moreover, the amplitude of cyclic relative variations in $\delta (LCA)$
 and $\delta (CRII)$ shows a similar latitudinal dependence (Fig.~\ref{Fig:6_scatter}b).
These results strongly favor the idea that the variations of LCA are related to
 CRII rather than other mechanisms, e.g., solar irradiance, which cannot naturally explain such
 a latitudinal dependence.

\section{Conclusions}

While in earlier studies data from a single neutron monitor was used as a proxy of cosmic
 ray intensity, we have explored the quantitative relationship between temporal and spatial
 variations of LCA and CRII over the globe for the period 1984--2000.
We suggest that the LCA time series can be decomposed
 into a long-term slow trend and inter-annual variations, the latter depicting a
 clear 11-year cycle in phase with CRII.
The trend whose nature is beyond the scope of the present study, is strong in
 tropical regions and possibly masks the LCA-CRII relation.
We then find a highly significant correlation between the de\-trended inter-annual LCA variations and CRII
 over the globe (polar regions being excluded).
A quantitative regression model was obtained (Eq.~\ref{Eq:lin2}), which
 implies a one-to-one relation between the relative variations of LCA and CRII over
 the latitude range 20--55$^\circ$S and  10--70$^\circ$N.
The amplitude of relative variations in LCA was found to increase polewards,
 in accordance with the amplitude of CRII variations but in
 contrast to the insolation which decreases polewards.
These results thus support the idea that LCA is modulated by CRII, rather
 than by solar irradiance, at inter-annual timescales between 1984-2000.

\begin{acknowledgments}
We  thank  Henrik Svensmark  for  useful  discussions.  The  financial
support by the Academy of Finland is acknowledged. OGG and
GAK  were  partly supported by the program  "Non-stationary
Processes in  Astronomy". The  cloud data are  obtained from  ISCCP D2
web-site (http://isccp.giss.nasa.gov/products/browsed2.html).
\end{acknowledgments}

\end{article}
\end{document}